\documentclass[twocolumn,showpacs,preprintnumbers,amsmath,amssymb,prl,aps]{revtex4-1}
\usepackage{graphicx}
\usepackage{bm}
\usepackage{float} 
\usepackage{amsmath}
\usepackage{gensymb}
\usepackage{epstopdf}
\usepackage[usenames, dvipsnames]{color}

\begin{document}

\title{Exceptional Dirac states in a non-centrosymmetric superconductor, BiPd}

\author{Arindam Pramanik$^1$}
\author{Ram Prakash Pandeya$^1$}
\author{D. V. Vyalikh$^{2,3}$}
\author{Alexander Generalov$^4$}
\author{Paolo Moras$^5$}
\author{Asish K. Kundu$^5$}
\author{Polina M. Sheverdyaeva$^5$}
\author{Carlo Carbone$^5$}
\author{Bhanu Joshi$^1$}
\author{A. Thamizhavel$^1$}
\author{S. Ramakrishnan$^1$}
\author{Kalobaran Maiti$^1$}
\altaffiliation{Corresponding author: kbmaiti@tifr.res.in}

\affiliation{$^1$Department of Condensed Matter Physics and Materials Science, Tata Institute of Fundamental Research, Homi Bhabha Road, Colaba, Mumbai - 400005, India \\
$^2$ Donostia International Physics Center (DIPC), 20018 Donostia San Sebasti\'{a}n, Basque Country, Spain\\
$^3$ IKERBASQUE, Basque Foundation for Science, 48013, Bilbao, Spain \\
$^4$ MAX IV Laboratory, Lund University, PO Box 118, 22100, Lund, Sweden\\
$^5$Istituto di Struttura della Materia, Consiglio Nazionale delle Ricerche, I-34149
Trieste, Italy}


\begin{abstract}
Quantum materials having Dirac fermions in conjunction with superconductivity is believed to be the candidate materials to realize exotic physics as well as advanced technology. Angle resolved photoemission spectroscopy (ARPES), a direct probe of the electronic structure, has been extensively used to study these materials. However, experiments often exhibit conflicting results on dimensionality and momentum of the Dirac Fermions (e.g. Dirac states in BiPd, a novel non-centrosymmetric superconductor), which is crucial for the determination of the symmetry, time-reversal invariant momenta and other emerging properties. Employing high-resolution ARPES at varied conditions, we demonstrated a methodology to identify the location of the Dirac node accurately and discover that the deviation from two-dimensionality of the Dirac states in BiPd proposed earlier is not a material property. These results helped to reveal the topology of the anisotropy of the Dirac states accurately. We have constructed a model Hamiltonian considering higher-order spin-orbit terms and demonstrate that this model provides an excellent description of the observed anisotropy. Intriguing features of the Dirac states in a non-centrosymmetric superconductor revealed in this study expected to have significant implication in the properties of topological superconductors.
\end{abstract}

\maketitle

\section{Introduction}

Recent times has seen the emergence of a new class of insulating materials, which are topological in nature. While the bulk of these materials is insulating, surface harbours partially filled (metallic) spin-split two dimensional bands with cone like structure (Dirac cone) arising due to the topological nature of the bulk bands. Bi$_2$Se$_3$ is one of the most studied materials in this category\cite{Shoucheng}, where the surface states and its evolution with impurities have been studied extensively \cite{Bi2Se3-massiveDF,Bi2Se3-O2,BieSe3-Deep}. The pool of topological materials have been enriched via discovery of Dirac Fermions as the surface states in superconductors such as BiPd \cite{Benia,Thirupati,Neupane}, $\beta$-PdBi$_2$ \cite{Sakano}, Cu$_x$Bi$_2$Se$_3$ \cite{Wray}, Sr$_x$Bi$_2$Se$_3$ \cite{Patnaik,Han}, etc. Among these topological materials, BiPd grabbed much attention as it stabilizes in noncentrosymmetric monoclinic structure($P2_1$) known as $\alpha$-BiPd and superconductivity appears below 3.8 K\cite{Joshi,Mondal,Sun}. Above 483 K, it undergoes polymorphic transition from $\alpha$-BiPd to orthorhombic $\beta$-BiPd (space group - $Cmc2_1$).

Due to the absence of inversion symmetry, (010) and (0$\bar{1}$0) faces of BiPd are inequivalent and the binding energy at the Dirac nodes on respective faces are also different. Interestingly, the twinning in the samples allows photoemission experiments to capture properties of both the surfaces simultaneously; while the Dirac bands on (010) face appear clearly in the experimental spectra, bands on (0$\bar{1}$0) face are often weak and appear in the immediate vicinity of the bulk states. Benia \textit{et al.}\cite{Benia} pointed out that the Dirac states in BiPd may not have topological origin as these are found in density functional calculations both with and without spin-orbit (SO) coupling. On the other hand, spin-resolved photoemission measurements have confirmed the spin-polarization of these states, which is a signature of topological behavior\cite{Neupane}. It is to note here that spin-polarized surface states are also observed in systems with heavy elements due to strong Rashba coupling.

Despite several studies, even the identification of the location of Dirac node and the dimensionality of the Dirac states are outstanding issues. Thirupathaiah \textit{et al.}\cite{Thirupati} reported this band to be found at $\overline{\Gamma}$ (Brillouin zone center). However, Yaresko \textit{et al.} have shown the Dirac states to be positioned at $\overline{S}$, a high-symmetry point at the surface Brillouin zone boundary based on their detailed density functional theoretical (DFT) calculations \cite{Yaresko}. In addition to this conflicting results on location of the Dirac node, Thirupathaiah \textit{et al.}\cite{Thirupati} proposed three dimensional nature of the Dirac states depicted by an energy gap at the Dirac node varying with $k_z$; although the repetitive nature of the gap as a function of $k_z$ was not observed. However, the DFT results characterize Dirac states as two dimensional surface states.\cite{Benia,Neupane} We carried out high-resolution angle-resolved photoemission spectroscopic (ARPES) measurements at carefully chosen experimental conditions and discover that the Dirac states are truly two dimensional; the anomalies reported earlier arose due to the sample alignment used in those experiments. Furthermore, we find that the anisotropy in the dispersion of the Dirac bands reported earlier \cite{Benia,Thirupati} appears far away from the Dirac node. We have constructed a model Hamiltonian considering higher-order spin-orbit terms, which provides a good description of all the features of the Dirac bands observed experimentally.

\section{experiment}

High-quality single crystals of BiPd were grown using modified Bridgman method. Crystal structure of the sample was determined via analysis of powder $x$-ray diffraction pattern and good crystallinity has been ensured employing Laue diffraction experiments. The lattice parameters found in the study correspond to the monoclinic structure as reported elsewhere.\cite{Joshi} Magnetization measurements exhibit a superconducting transition at 3.8 K. ARPES measurements were performed at Diamond Light Source, United Kingdom and Elettra, Italy. At Diamond Light Source, experiments were carried out at I05 beamline \cite{I05beamline} at a temperature of 10 K, base pressure of 5$\times$10$^{-11}$ torr and energy resolution of 5 meV. Measurements at Elettra were done at VUV beamline at a temperature of 25 K, base pressure of 6$\times$10$^{-11}$ torr and energy resolution of 10 meV.

\section{Results and Discussions}

\begin{figure}
\includegraphics [scale=0.4] {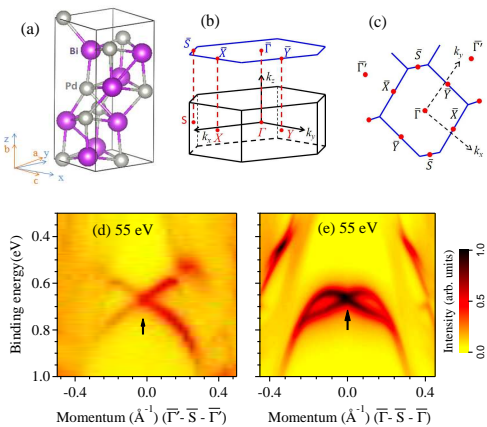}
 \caption{(a) Unit cell of BiPd in real space. Angles between $c$ \& $x$-axis and $a$ \& $y$-axis are close to 5.5$^o$. (b) Bulk and (c) surface Brillouin zones. ARPES data (d) along $\overline{\Gamma^\prime}-\overline{S}-\overline{\Gamma^\prime}$ and (e) $\overline{\Gamma}-\overline{S}-\overline{\Gamma}$ vectors. Dirac point is identified with an arrow and the $k$-axis is shifted to make $\overline{S}$-point as zero.}
 \label{set1}
\end{figure}

\begin{figure}
\includegraphics [scale=0.4] {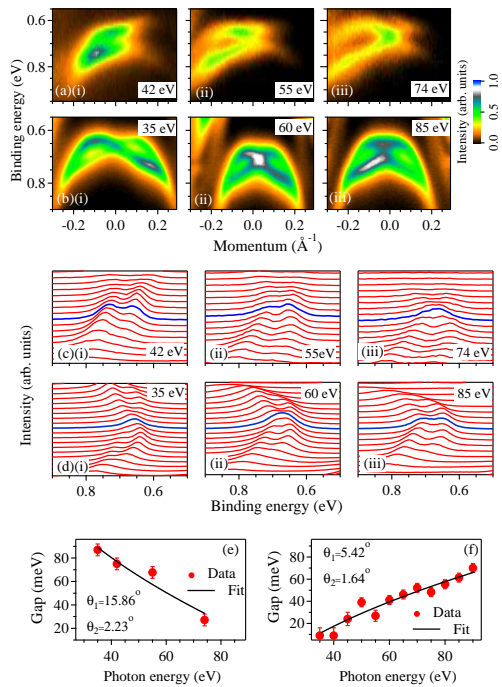}
\caption{(a)(i)-(iii) First set of ARPES data collected along $\overline{\Gamma}-\overline{S}-\overline{\Gamma}$ direction after normalizing the sample using 74 eV photon energy. (b)(i)-(iii) Similar second set of ARPES data collected after sample normalization using 35 eV photon energy. Here, sample normalization means, fixing the polar angle such that there is no gap at the Dirac node. (c)(i)-(iii) and (d)(i)-(iii) show the EDCs derived from the first and second set of the spectra. Blue lines mark the EDCs passing through the Dirac cone apex. The energy gaps derived at the Dirac cone apex at $\overline{S}$-point from (e) the first and (f) the second set of the spectra. Photon energy is used as independent variable for both the plots. Black lines superimposed over the data points represent the estimated energy gaps considering the sample alignment angles, $\theta_1$ (azimuthal) and $\theta_2$ (polar) as defined in Fig. \ref{set3}. Derived values of the angles are also shown.}
 \label{set2}
\end{figure}

The crystal unit cell of BiPd and the crystal axes defined in the cartesian coordinate system are shown in Fig.~\ref{set1}(a). The $a$ and $c$-axis make almost equal angle of about 5.5$^o$ with $x$- and $y$-axis, respectively. The $z$-axis is defined along the crystal axis, $b$. In Figs. ~\ref{set1}(b) and \ref{set1}(c), we show the bulk Brillouin zone (BZ) and its projection on the surface, respectively. Photoemission spectra along two directions are shown in Figs.~\ref{set1}(d) and \ref{set1}(e) with Dirac states appearing at 0.66 eV binding energy shown using an arrow. Our $k$-space analysis of the data as discussed in Fig. \ref{set2} suggests that Dirac node is positioned at $S$ point in the surface Brillouin zone, which is about 0.73 \AA$^{-1}$ away from the $\overline{\Gamma}$ point; in the figure, the $k$-axis is shifted to make it zero at the Dirac node. While the bands along $\overline{\Gamma^\prime}-\overline{S}-\overline{\Gamma^\prime}$ exhibit spin splitting varying monotonically with momentum, the momentum dependence of spin-splitting along $\overline{\Gamma}-\overline{S}-\overline{\Gamma}$ is non-linear; it increases to a maximum near momentum of $0.1\AA^{-1}$ away from $S$ and then decreases.

To investigate the dimensionality of the Dirac states, we acquired spectra at various photon energies, which helps to decide surface or bulk nature of the bands. In Figs.~\ref{set2}(a)(i)-(iii), we show a set of spectra along with corresponding energy distribution curves (EDC) in Figs.~\ref{set2}(c)(i)-(iii). Second set of spectra is displayed in Figs.~\ref{set2}(b)(i)-(iii) with corresponding EDCs in Figs.~\ref{set2}(d)(i)-(iii). Pieces cut from the same single crystal were used to obtain these two sets of spectra. A close inspection reveals interesting differences between the two sets. In the first set (sample position optimized using 74 eV photon energy), as the photon energy is lowered, the Dirac cone becomes more asymmetric and the bands do not cross each other. The energy gap at the Dirac node was derived by fitting two peaks in the EDC (blue curves in Figs.~\ref{set2}(c)) across the node. The gap increases as the probing energy is lowered. The second set of spectra were collected after optimizing the sample position at 35 eV photon energy so that the Dirac node is captured. Curiously, the second set exhibits an opposite trend of the gap at the Dirac node; the energy gap increases with the increase in photon energy as manifested clearly in the EDCs (blue curves in Fig.~\ref{set2}(d)). In this case, the dispersion of the Dirac bands remains symmetric over the energy range studied. In Figs. \ref{set2}(e) \& \ref{set2}(f), we show the derived energy gaps of the first and second sets, respectively exhibiting opposite trend. If the Dirac node appears at the $\overline{\Gamma}$-point, such a scenario cannot arise.

\begin{figure}
\includegraphics [scale=0.4] {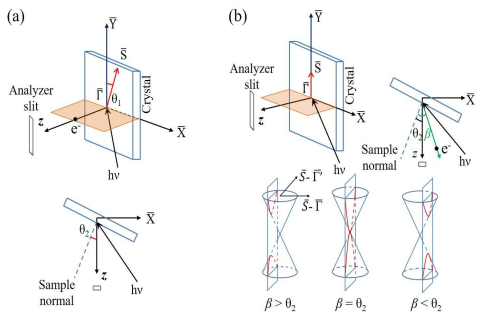}
\caption{Schematics of the experimental geometry. (a) The geometry used in the present study where $\theta_1$ and $\theta_2$ are in-plane (azimuthal) and out-of-plane (polar) angles, respectively. (b) A special case where the in-plane angle, $\theta_1 = 90^o$. The angle, $\beta$ is the emission angle of photoelectrons corresponding to the Dirac node with respect to the sample surface-normal. Conic sections in the lower panel show the expected band dispersions in the spectra for different values of $\beta$ with respect to $\theta_2$ due to the change in photon energy keeping the sample orientation unchanged.}
 \label{set3}
\end{figure}

To address this perplexing scenario, we investigate the experimental geometry carefully and define an axis system as given in Fig. \ref{set3}(a). The angle, $\theta_1$ is defined as the azimuthal angle made by the vector, $\overline{\Gamma}-\overline{S}$ with the analyzer slit, which is aligned along $y$ direction. The tilt angle is labeled as $\theta_2$; the angle of the analyser with the sample-surface-normal. Depending on the magnitudes of $\theta_1$ and $\theta_2$, different trends of the energy gap at the Dirac node is expected as a function of probing photon energy. To digress for a moment, we demonstrate a less complex scenario in Fig.~\ref{set3}(b); $\theta_1$ = 90$^o$ \& $\theta_2\neq$~0. Here, $\overline{\Gamma^\prime}-\overline{S}-\overline{\Gamma^\prime}$ vector lies along the slit (probed $k$-vector). If the Dirac node is located at a finite momentum along the $\overline{\Gamma}-\overline{S}$ direction, corresponding electrons will emerge at an angle, $\beta$ with respect to the sample-normal which is dependent on the photon energy used for experiments; with the increase of the photon energy, $\beta$ gradually reduces. For sufficiently low photon energy, $\beta$ will be larger than $\theta_2$. It becomes equal to $\theta_2$ at some photon energy and then becomes smaller for higher photon energies. The lower panel of Fig.~\ref{set3}(b) depicts the schematics of the acquired dispersion using conic section. The vertical plane is the plane of constant momentum along $\overline{\Gamma}-\overline{S}$, which lies parallel to the analyzer slit. Intersection of this vertical plane and the cone determines the shape of the dispersion (red curves) as seen in the spectra. When $\beta \neq \theta_2$, instead of a cone the Dirac state manifests itself as two hyperbolas separated by an energy gap. An ideal cone with a Dirac node is imaged at a particular photon energy when the corresponding $\beta$ becomes equal to $\theta_2$.

Now, we investigate the spectra presented in Fig.~\ref{set2}; the presence of finite $\theta_1$ will manifest as an asymmetry in the cone structure as the probed $k$-vector does not pass through the Dirac node. This argument is verified from the values of $\theta_1$ obtained from fittings in Figs.~\ref{set2}(e)\& \ref{set2}(f); the first set of spectra (see Fig.~\ref{set2}(a)) exhibit strong asymmetry and correspond to a higher value of $\theta_1$ compared to the second set shown in Fig.~\ref{set2}(b). Derived values of $\theta_2$ are also listed in the figures. Excellent representation of the experimental results establishes that the gap at the Dirac node is not the property of the material but arises due to the sample alignment. Zero gap within the experimental error can be obtained for both sets of data at all probing energies once the fitted curve is subtracted from the measured gap. This suggests that opening of an energy gap at the node is not an inherent property of the system. We have verified this experimentally; we observed that the sample realignment at different photon energies leads to reduction of the energy gap to zero. For example, see the data in Figs. \ref{set1}(d) and \ref{set1}(e) exhibiting distinct Dirac node for photon energy of 55 eV although the spectra collected using other setup for the same photon energy show non-zero gap. This is in addition to the data in Figs. \ref{set2}(a)(iii) and \ref{set2}(b)(i) for 74 eV and 35 eV, respectively. This establishes the finite momentum at the Dirac node and two dimensional nature of the Dirac states as there is no observable variation of the bands with the photon energy. Using this scenario, earlier ARPES results can be captured excellently well; as a representative case, we analysed ARPES data of Thirupathaiah {\it et al.} \cite{Thirupati} exhibiting identical behavior for $\theta_2$ = 9.6$^o$ and $\theta_1$ = 90$^o$.

\begin{figure}
\includegraphics [scale=0.4] {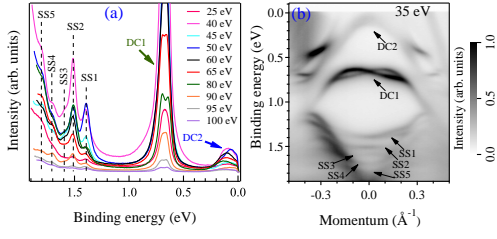}
\caption{(a) EDCs passing through the Dirac node are plotted for multiple photon energies ranging from 25 eV to 100 eV. DC1 and DC2 are the Dirac cones on $010$ and $0\bar{1}0$ surfaces, respectively. DC1 exhibits splitting into two peaks where the alignment of the sample is away from $\overline{S}$-point due to the change in photon energy. SS1-SS5 label the surface states at higher binding energies. (b) A representative spectrum at 35 eV shows all the observed surface states of BiPd.}
 \label{set4}
\end{figure}

Besides the widely discussed Dirac states, we discover few more two-dimensional states lying at higher binding energies. In Fig.~\ref{set4}(a), we show the EDCs at various photon energies taken across the Dirac node. DC1 is the Dirac state under investigation. DC2 shown in the inset of Fig. \ref{set4}(a) is the Dirac cone on (0$\bar{1}$0) surface \cite{Benia}. In addition, few other states labeled as SS1-SS5 are seen to be positioned at fixed binding energy even as the photon energy is varied over a large range.

\begin{figure}
\includegraphics [scale=0.4] {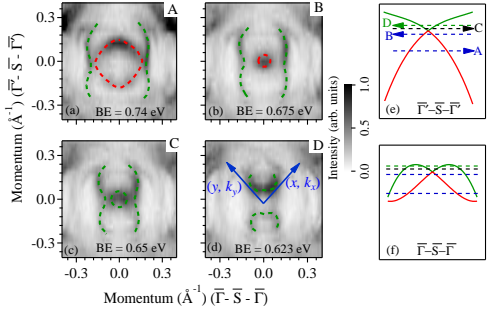}
\caption{Constant-energy contours in the $xy$-plane at binding energies (a) 0.74 eV, (b) 0.675 eV, (c) 0.65 eV and (d) 0.623 eV obtained from the ARPES data at 55 eV photon energy as shown in Fig. \ref{set1}. The binding energy positions of the contours are labeled as A, B, C and D and shown schematically in (e) along the $k$-vector, $\overline{\Gamma^\prime}-\overline{S}-\overline{\Gamma^\prime}$ and (f) along the $k$-vector, $\overline{\Gamma}-\overline{S}-\overline{\Gamma}$. Green and red colors identify the contours derived from green and red colored energy bands. Cartesian coordinate system is shown in (d); ($x$,$k_x$) lies at $45^{o}$ with respect to $\overline{\Gamma}-\overline{S}-\overline{\Gamma}$.}
 \label{set5}
\end{figure}

We now address the issue of anisotropy of the Dirac bands; such anisotropy has also been reported in other materials. For example, Bi$_2$Te$_3$\cite{Chen, Souma} and Bi$_2$Se$_3$\cite{Kuroda, Valla} are two prominent cases of this class. There are other cases too, such as Ru$_2$Sn$_3$\cite{Gibson}, $\beta$-Bi$_4$I$_4$\cite{Autes}, $\beta$-HgS\cite{Virot}, $\beta$-Ag$_2$Te \cite{Zha}, Au film grown on Ag(111)\cite{Requist}, Ag film grown on Au (111)\cite{Requist} etc. Anisotropy in these systems is attributed to the symmetries at the surfaces. In Figs.~\ref{set5}(a) - \ref{set5}(d), we show the constant-energy contours of BiPd taken across the Dirac states. Energy positions of the contours are shown using schematics in Figs.~\ref{set5}(e) and \ref{set5}(f) with dashed lines. Each constant-energy map consists of two contours. Green and red colors are used to identify the contours with the energy bands. Shape of the outer contour (cut A, B \& C) exhibits the twofold rotational symmetry of the crystal belonging to $C_2$ point group. Interestingly, the inner contour near the nodal point is isotropic (cut B \& C) and gradually evolves into a twofold symmetric curve (cut A) far away from the nodal point. At the top of the Dirac cone above the node, the contour evolves into two disjoint segments (cut D). Further, we notice that the top and bottom portions of the outer contour are missing in all the cuts including D. This picture is consistent with the observation in Fig.~\ref{set1}(d), where along $\overline{\Gamma^\prime}-\overline{S}-\overline{\Gamma^\prime}$, the surface bands merge with the bulk band resulting into an incomplete contour. Signature of twofold rotational symmetry in contours implies that electrons are subject to the effects of crystal potential. Benia \textit{et al.} argued that a twofold rotational symmetry in Rashba coupling strength at the surface is responsible for this anisotropic band dispersion \cite{Benia}. However, it is impossible to capture the non-parabolic dispersion along $\overline{\Gamma}-\overline{S}-\overline{\Gamma}$ (Fig.~\ref{set1} (e)) as well as the contour shapes shown in Fig.~\ref{set5} using this scenario. Moreover, a $C_2$ symmetric Rashba coupling, which is first order in momentum ($k_x, k_y$), will produce two fold symmetric contours near the node unlike the circular contours seen here (cut B \& C).

\begin{figure}
\includegraphics [scale=0.4] {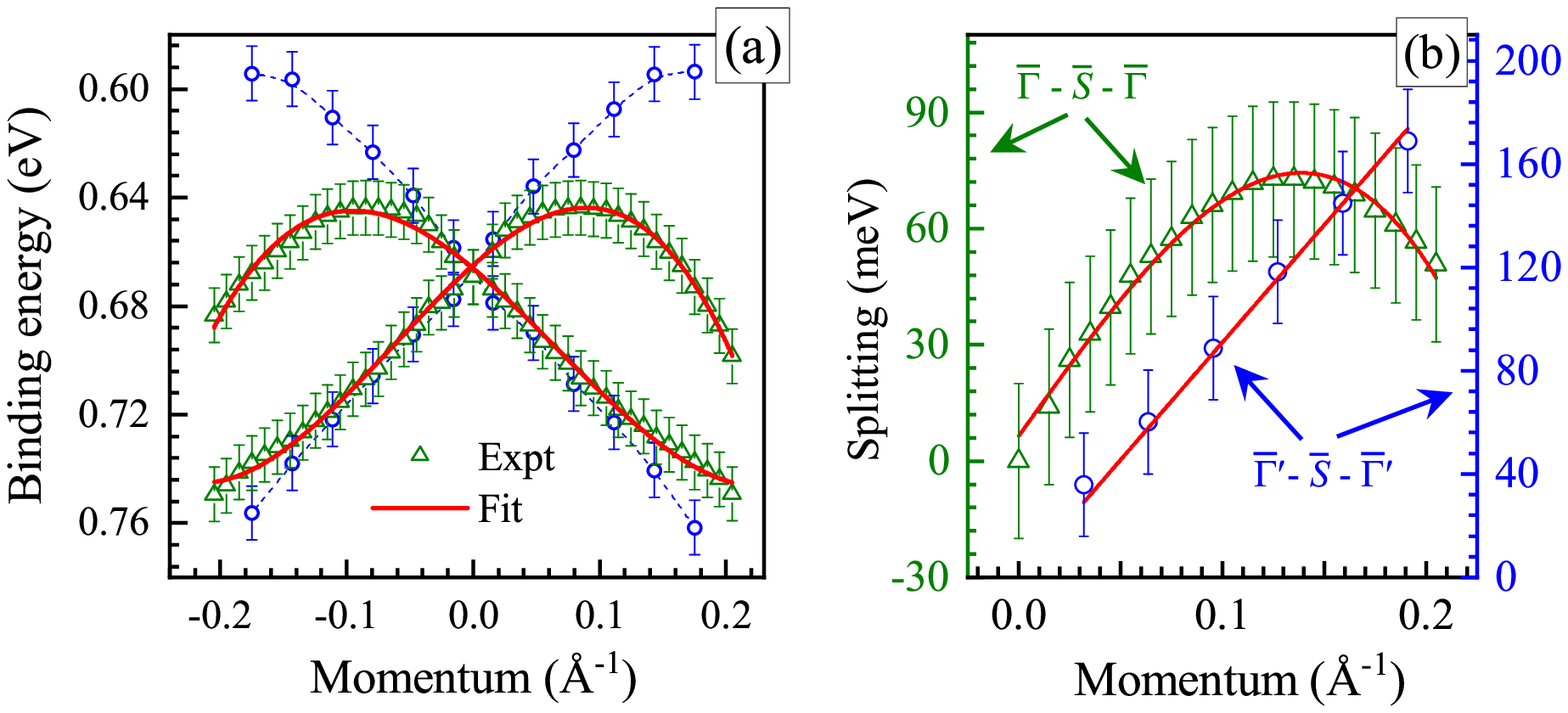}
\vspace{-36ex}
\caption{(a) Dispersion along $\overline{\Gamma}^\prime-\overline{S}-\overline{\Gamma}^\prime$ (blue open circles) and $\overline{\Gamma}-\overline{S}-\overline{\Gamma}$ (green open triangles) derived from EDCs. Red line superimposed over the triangles represent the fit based on the eigenvalue equation \ref{ham1}. (b) Spin splitting along $\overline{\Gamma}-\overline{S}-\overline{\Gamma}$ (green open triangles) and $\overline{\Gamma^\prime}-\overline{S}-\overline{\Gamma^\prime}$ (blue open circles). Lines represent the fits based on the eigenvalue equation.~\ref{ham1}.}
 \label{set6}
\end{figure}

Experimentally observed dispersions of the Dirac bands along two orthogonal directions are shown in Fig.~\ref{set6}(a). It is evident in the figure that although the dispersions are different far away from the node, they are identical near the Dirac node. This again confirms the isotropy near the nodal point as found in constant energy contours of Fig.~\ref{set5}. The observation of circular contours near the node, gradual loss of isotropy away from the node and the non-parabolic nature of the dispersion along $\overline{\Gamma^\prime}-\overline{S}-\overline{\Gamma^\prime}$ together suggest significant contributions of higher order spin-orbit terms \cite{Liang, Liu}. Considering the $C_2$ point group of the material, we derive the Hamiltonian up to the third order term in momentum. Under $C_2$ symmetry, $H(\vec{k})$ is required to remain invariant, i.e. $C_2 H(\vec{k})C_2^{-1}= H(\vec{k})$. Time reversal symmetry puts additional constraint, $H(\vec{k})= \Theta H(-\vec{k})\Theta^{-1}=\sigma_y H^*(-\vec{k})\sigma_y$. The choice of the crystal axes are shown in Fig.~\ref{set1}(a)/\ref{set5}(d). $x$-axis is chosen to lie at $45^o$ with respect to $\overline{\Gamma}-\overline{S}-\overline{\Gamma}$ direction. A model Hamiltonian with the $\overline{S}$ point as reference point is constructed as follows.

$$H = E_0+\frac{\hbar^2 k_x^2}{2m_x^*}+\frac{\hbar^2 k_y^2}{2m_y^*}+\theta k_xk_y+\alpha (\sigma_x k_y - \sigma_y k_x)$$
$$+\gamma(\sigma_xk_x-\sigma_yk_y)+\delta(\sigma_xk_x+\sigma_yk_y)$$
$$+q(\sigma_xk_xk_y^2+\sigma_yk_yk_x^2)+r (\sigma_x k_x k_y^2 - \sigma_y k_y k_x^2)$$
\begin{equation}
+ s(\sigma_y k_x k_y^2 + \sigma_x k_y k_x^2)+ t (\sigma_y k_x k_y^2 - \sigma_x k_y k_x^2)
\label{ham1}
\end{equation}

$m_x^*$, $k_x$ and $m_y^*$, $k_y$ are effective mass of electron and momentum along the $x$ and $y$ axis, respectively. Due to isotropy near the $\overline{S}$ point, effective mass along the two axes are set to have equal value, $m^*$. $\sigma_x$ and $\sigma_y$ are the Pauli matrices in the spin space; $E_0$, $\theta$, $\alpha$, $\gamma$, $\delta$, $q$, $r$, $s$ and $t$ are the parameters of the Hamiltonian.

Band dispersion and spin splitting extracted from EDCs from the spectrum at 55 eV are shown by symbols in Fig.~\ref{set6}(a) and \ref{set6}(b), respectively. Spin splitting along $\overline{S}-\overline{\Gamma^\prime}$ direction ($k_x = k_y$) is represented by open circles and along $\overline{S}-\overline{\Gamma}$ direction ($k_x = -k_y$), it is open triangles. The spin splitting along $\overline{S}-\overline{\Gamma^\prime}$ varies linearly with momentum as expected for Rashba split parabolic bands. Along $\overline{S}-\overline{\Gamma}$, away from $\overline{S}$ point, the bands deviate from the expected parabolic dispersion. Consequently, only close to $\overline{S}$ point splitting varies linearly with momentum, while at large momentum effect of higher order takes over. Fitting of all the curves are done using expressions derived from equation~\ref{ham1}. In each case, good quality of fitting is achieved for momentum region close to $\overline{S}$. The parameters for the fitting shown in the figure are given in Table I. Large values of $r$ (Dresselhaus term) and $t$ suggest importance of the higher-order terms to capture the experimental scenario. It appears that the absence of inversion symmetry plays an important role in the spin-orbit coupling as also manifested in the core level spectra \cite{BiPD-HXPS}. It is to note here that $\alpha$ and $\gamma$ both cannot be nonzero simultaneously because it breaks the isotropy near $\overline{S}$ point \cite{Gani1,Gani2}. Hence, when $\alpha (\gamma)$ is nonzero $\gamma(\alpha)$ becomes zero. This gives us two possible cases, $H(\alpha\neq0, \gamma=0)$ and $H(\alpha=0,\gamma\neq0)$ for the bands under consideration (discussed later).

\begin{table}
\caption{Obtained values of the parameters (units: eV, \AA).}
\begin{ruledtabular}
\begin{tabular}{|l|c|c|c|c|c|c|c|c|c|}
Parameter & $m^*$ & $E_0$ & $\theta$ & $\alpha (-\gamma)$ & $\delta$ & $q$ & $r$ & $s$ & $t$\\
\hline
Value & -2.6$m_e$ & -0.664 & 0 & 0.37 & 0 & 0 & 6.64 & 0 & 6.64 \\
\end{tabular}
\end{ruledtabular}
\end{table}

\begin{figure}
\includegraphics [scale=0.4] {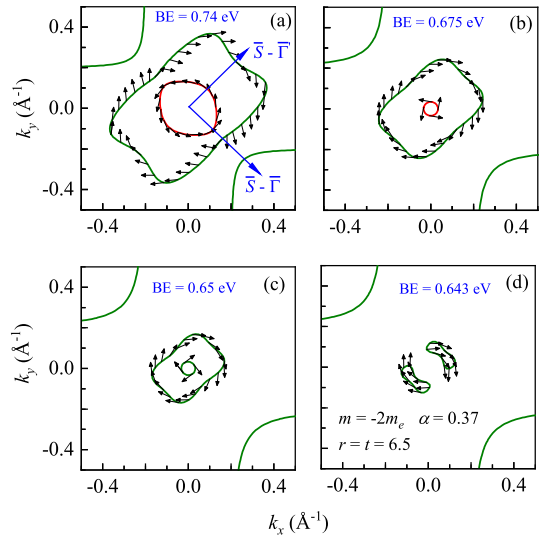}
 \vspace{-18ex}
\caption{Contours simulated at (a) 0.74 eV, (b) 0.675 eV, (c) 0.65 eV and (d) 0.643 eV binding energies using the Hamiltonian in equation.~\ref{ham1}. Parameters used for the simulations are listed in (d). Colors (red or green) of contours are in accordance with Fig.~\ref{set5}. Black arrows show the directions of the spin polarization.}
 \label{set7}
\end{figure}

To confirm that obtained parameters can reproduce the experimental results shown in Fig.~\ref{set5}, we simulate constant energy maps with values of parameters similar to obtained values. Both the Hamiltonians, $H(\alpha=0.37, \gamma=0)$ and $H(\alpha=0, \gamma=-0.37)$, produce identical contours. In Fig.~\ref{set7}, we show the simulated contours at binding energies similar to that shown in Fig.~\ref{set5}. Note that since the size of the model Hamiltonian is 2$\times$2, it describes only the spin-split Dirac like states and does not capture the merger of the Dirac states with the bulk bands along $\overline{S}-\overline{\Gamma^\prime}$ direction. Hence, it is unable to reproduce incomplete shape of the outer contour; instead it produces closed curves. Evidently, the shapes of the contours provide an excellent description of the experimentally observed scenario around $\overline{S}$-point.

We now study the spin-texture of the concerned terms in the Hamiltonian to identify the correct form. Fitting of dispersion alone cannot decide whether the linear term with coefficient $\alpha$ or $\gamma$ appears in the Hamiltonian; both of these produce identical contours. However, these two possibilities give rise to two different spin-textures of the states which can be determined by making two dimensional vector plot of the effective magnetic field arising due to the individual terms. $\alpha$ term produces helical spin texture \cite{Gani1, Gani2}. Presence of nonzero electric dipole moment is required for this spin texture to occur in the systems. This term is responsible for spin-splitting at the surface due to nonzero electric field (Rashba effect \cite{Rashba}). Whereas $\gamma$ term gives rise to Dresselhaus spin texture\cite{Gani1,Gani2}, which arises due to absence of inversion symmetry in the bulk \cite{Dressel}. Evidently, Dirac states in BiPd, being two-dimensional surface states, will assume a helical spin-texture and $H(\alpha=0.37, \gamma=0)$ will be relevant for the present scenario. We note here that unlike in some systems (Bi$_2$Se$_3$ \cite{Liang,Nomura}, Bi$_2$Te$_3$ \cite{Liang,Souma}, $\beta$-Ag$_2$Te \cite{Zha} etc.), where higher order spin orbit coupling gives rise to out-of-plane spin polarization, surface states in BiPd will not possess out-of-plane spin component. Spin splitting term containing $\sigma_z$, which leads to out-of-plane spin polarization, is absent in the Hamiltonian (see equation.~\ref{ham1}) owing to the $C_2$ symmetry of the system. Arrows in the Fig.~\ref{set7} exhibit the direction of spin polarization along the contours, which are in-plane\cite{Usachov}.

\section{Conclusions}

In summary, we studied the Dirac states in a non-cetrosymmetric superconductor, BiPd. High-quality of the sample and high-resolution of the ARPES technique employed in this study helped to reveal subtle features in the electronic structure. Our experimental results helped to identify the momentum of the Dirac node and establish the two dimensional character of the Dirac states resolving the outstanding disputes on these two issues. This study brings out the importance of deriving correct experimental geometry to reveal experimental results related to the properties of materials, in particular, the cases where the point of interest is not the center of the Brillouin zone ($\overline{\Gamma}$-point). This is crucial for the identification of the symmetry properties, time reversal invariant momenta and their implication in various other exoticity of the material. In addition, we discover several other surface states at binding energies higher than the Dirac point revealing complexity of the system. Since the Rashba term alone cannot adequately capture the experimental results, we constructed a model Hamiltonian including spin-orbit coupling terms of higher-order in momentum. Our model provide an excellent description of the anisotropy of the Dirac states. The necessity of the higher-order terms including Dresselhaus type terms reveals importance of the absence of inversion symmetry in the electronic properties of such systems.


\section{Acknowledgements}
Authors acknowledges financial support from the Department of Atomic Energy, Govt. of India under the project no. 12-R\&D-TFR-5.10-0100, Diamond Light Source for beamtime at the beamline I05 under proposal SI11512. DVV acknowledges support from the Spanish Ministry of Economy (No. MAT-2017-88374-P). KM acknowledges financial support from DAE, Govt. of India under the DAE-SRC-OI Award program, and Department of Science and Technology, Govt. of India under J. C. Bose Fellowship program.


\begin{thebibliography}{99}
%
\bibitem{Shoucheng}
H. Zhang, C. X. Liu, X. L. Qi, X. Dai, Z. Fang, and S. C. Zhang, \textit{Nat. Phys.} \textbf{5}, 438 (2009).
%
\bibitem{Bi2Se3-massiveDF}
Y. L. Chen, J. H. Chu, J. G. Analytis, Z. K. Liu, K. Igarashi, H. H. Kuo, X. L. Qi, S. K. Mo, R. G. Moore, D. H. Lu, M. Hashimoto, T. Sasagawa, S. C. Zhang, I. R. Fisher, Z. Hussain, and Z. X. Shen,  \textit{Science} {\bf 329}, 659 (2010).
%
\bibitem{Bi2Se3-O2}
D. Kong, J. J. Cha, K. Lai, H. Peng, J. G. Analytis, S. Meister, Y. Chen, H. J. Zhang, I. R. Fisher, Z. X. Shen, and Y. Cui, \textit{ACS Nano} {\bf 5}, 4698–4703 (2011).
%
\bibitem{BieSe3-Deep}
D. Biswas, S. Thakur, K. Ali, G. Balakrishnan, and K. Maiti, \textit{Sci. Rep.} {\bf 5}, 10260 (2015); D. Biswas, S. Thakur, G. Balakrishnan, and K. Maiti, \textit{Sci. Rep.} {\bf 5}, 17351 (2015); D. Biswas and K. Maiti, \textit{EPL} {\bf 110}, 17001 (2015); D. Biswas and K. Maiti, \textit{J. Elec. Spec. Relat. Phenom.} {\bf 208}, 90 (2016).
%
\bibitem{Benia}
H. M. Benia, E. Rampi, C. Trainer, C. M. Yim, A. Maldonado, D. C. Peets, A. St\"{o}hr, U. Starke, K. Kern, A. Yaresko, G. Levy, A. Damascelli, C. R. Ast, A. P. Schnyder, and P. Wahl, \textit{Phys. Rev. B} \textbf{94}, 121407(R) (2016).
%
\bibitem{Thirupati}
S. Thirupathaiah, S. Ghosh, R. Jha, E.D.L. Rienks, K. Dolui, V.V.R. Kishore, B. B\"{u}chner, T. Das, 
V.P.S. Awana, D.D. Sarma, and J. Fink, \textit{Phys. Rev. Lett.} \textbf{117}, 177001 (2016).
%
\bibitem{Neupane}
M. Neupane, N. Alidoust, M. M. Hosen, J. X. Zhu, K. Dimitri, S. Y. Xu, N. Dhakal, R. Sankar, I. Belopolski, D. S. Sanchez, T. R. Chang, H. T. Jeng, K. Miyamoto, T. Okuda, H. Lin, A. Bansil, D. Kaczorowski, F. Chou, M. Zahid Hasan, and T. Durakiewicz, \textit{Nat. Commun.} \textbf{7}, 13315 (2016).
%
\bibitem{Sakano}
M. Sakano, K. Okawa, M. Kanou, H. Sanjo, T. Okuda, T. Sasagawa, and  K. Ishizaka, \textit{Nat. Commun.} \textbf{6}, 8595 (2015).
%
\bibitem{Wray}
L. A. Wray, S. Y. Xu, Y. Xia, Y. S. Hor, D. Qian, A. V. Fedorov, H. Lin, A. Bansil, R. J. Cava, and M. Z. Hasan, \textit{Nat. Phys.} \textbf{6}, 855-859 (2010).
%
\bibitem{Patnaik}
Shruti, V. K. Maurya, P. Neha, P. Srivastava, and S. Patnaik, \textit{Phys. Rev. B} {\bf 92}, 020506(R) (2015).
%
\bibitem{Han}
C. Q. Han, H. Li, W. J. Chen, F. Zhu, M. Y. Yao, Z. J. Li, M. Wang, B. F. Gao, D. D. Guan, C. Liu, C. L. Gao, D. Qian, and J. F. Jia, \textit{Appl. Phys. Lett.} \textbf{107}, 171602 (2015).
%
\bibitem{Joshi}
B. Joshi, A. Thamizhavel, and S. Ramakrishnan, \textit{Phys. Rev. B} \textbf{84}, 064518 (2011).
%
\bibitem{Mondal}
M. Mondal, B. Joshi, S. Kumar, A. Kamlapure, S. C. Ganguli, A. Thamizhavel, S. S. Mandal, S. Ramakrishnan, and P. Raychaudhuri, \textit{Phys. Rev. B} \textbf{86}, 094520 (2012).
%
\bibitem{Sun}
Z. Sun,  M. Enayat, A. Maldonado, C. Lithgow, E. Yelland, D. C. Peets, A. Yaresko, A. P. Schnyder, and P. Wahl, \textit{Nat. Commun.} \textbf{6}, 6633 (2015).
%
\bibitem{Yaresko}
A. Yaresko, A. P. Schnyder, H. M. Benia, C. M. Yim, G. Levy, A. Damascelli, C. R. Ast, D. C. Peets, and P. Wahl, \textit{Phys. Rev. B} \textbf{97}, 075108 (2018).
%
\bibitem{I05beamline}
 M. Hoesch,  T. K. Kim, P. Dudin,  H. Wang, S. Scott, P. Harris, S. Patel, M. Matthews, D. Hawkins,  S. G. Alcock, T. Richter, J. J. Mudd, M. Basham, L. Pratt, P. Leicester, E. C. Longhi,  A. Tamai, and F. Baumberger, \textit{Rev. Sci. Inst.} {\bf 88}, 013106 (2017).
%
\bibitem{Chen}
Y. L. Chen, J. G. Analytis, J. H. Chu, Z. K. Liu, S. K. Mo, X. L. Qi, H. J. Zhang, D. H. Lu, X. Dai, Z. Fang, S. C. Zhang, I. R. Fisher, Z. Hussain, and Z.-X. Shen, \textit{Science} \textbf{325}, 5937 (2009).
%
\bibitem{Souma}
S. Souma, K. Kosaka, T. Sato, M. Komatsu, A. Takayama, T. Takahashi, M. Kriener, K. Segawa, and Y. Ando, \textit{Phys. Rev. Lett.} \textbf{106}, 216803 (2011).
%
\bibitem{Kuroda}
K. Kuroda, M. Arita, K. Miyamoto, M. Ye, J. Jiang, A. Kimura, E. E. Krasovskii, E. V. Chulkov, H. Iwasawa, T. Okuda, K. Shimada, Y. Ueda, H. Namatame, and M. Taniguch, \textit{Phys. Rev. Lett.} \textbf{105}, 076802 (2010).
%
\bibitem{Valla}
T. Valla, Z. H. Pan, D. Gardner, Y. S. Lee, and S. Chu, \textit{Phys. Rev. Lett.} \textbf{108}, 117601 (2012).
%
\bibitem{Gibson}
Q. D. Gibson, D. Evtushinsky, A. N. Yaresko, V. B. Zabolotnyy, M. N. Ali, M. K. Fuccillo, J. Van den Brink, B. B$\ddot{\text{u}}$chner, R. J. Cava, and S. V. Borisenko, \textit{Sci. Rep.} \textbf{4}, 5168 (2014).
%
\bibitem{Autes}
G. Aut$\grave{\text{e}}$s, A. Isaeva3, L. Moreschini, J. C. Johannsen, A. Pisoni, R. Mori, W. Zhang, T. G. Filatova, A. N. Kuznetsov, L. Forr$\acute{\text{o}}$, W. V. den Broek, Y. Kim, K. S. Kim, A. Lanzara, J. D. Denlinger, E. Rotenberg, A. Bostwick, M. Grioni, and O. V. Yazyev, \textit{Nat. Mater.} \textbf{15}, 154 (2015).
%
\bibitem{Virot}
F. Virot, R. Hayn, M. Richter, and J. van den Brink, \textit{Phys. Rev. Lett.} \textbf{106}, 236806 (2011).
%
\bibitem{Zha}
W. Zhang, R. Yu, W. Feng, Y. Yao, H. Weng, Xi Dai, and Z. Fang, \textit{Phys. Rev. Lett.} \textbf{106}, 156808 (2011).
%
\bibitem{Requist}
R. Requist, P. M. Sheverdyaeva, P. Moras, S. K. Mahatha, C. Carbone, and E. Tosatti, \textit{Phys. Rev. B} \textbf{91}, 045432 (2015).
%
\bibitem{Liang}
L. Fu, \textit{ Phys. Rev. Lett.} \textbf{103}, 266801 (2009).
%
\bibitem{Liu}
C. X. Liu, X. L. Qi, H. Zhang, X. Dai, Z. Fang, and S. C. Zhang, \textit{ Phys. Rev. B} \textbf{82}, 045122 (2010).
%
\bibitem{BiPD-HXPS} 
A. Pramanik, R. P. Pandeya, K. Ali, B. Joshi, I. Sarkar, P. Moras, P. M. Sheverdyaeva, A. K. Kundu, C. Carbone, A. Thamizhavel, S. Ramakrishnan, and K. Maiti, \textit{Phys. Rev. B} {\bf 101}, 035426 (2020).
%
\bibitem{Gani1}
S. D. Ganichev, V. V. Bel\'{k}ov, L. E. Golub, E. L. Ivchenko, P. Schneider, S. Giglberger, J. Eroms, J. D. Boeck, G. Borghs, W. Wegscheider, D. Weiss, and W. Prettl, \textit{Phys. Rev. Lett.} \textbf{92}, 256601 (2004).
%
\bibitem{Gani2}
S. D. Ganichev and L. E. Golub, \textit{Phys. Status Solidi B} \textbf{251}, No. 9, 1801-1823 (2014).
%
\bibitem{Rashba}
Y. A. Bychkov and E. I. Rashba, \textit{JETP Lett.} \textbf{39}, 78-81 (1984).
%
\bibitem{Dressel}
G. Dresselhaus, \textit{Phys. Rev.} \textbf{100}, 580-6 (1955).
%
\bibitem{Nomura}
M. Nomura, S. Souma, A. Takayama, T. Sato, T. Takahashi, K. Eto, K. Segawa, and Y. Ando, \textit{Phys. Rev. B} \textbf{89}, 045134 (2014).
%
\bibitem{Usachov}
D. Yu. Usachov, I. A. Nechaev, G. Poelchen, M. Güttler, E. E. Krasovskii, S. Schulz, A. Generalov, K. Kliemt, A. Kraiker, C. Krellner, K. Kummer, S. Danzenbächer, C. Laubschat, A. P. Weber, J. Sánchez-Barriga, E. V. Chulkov, A. F. Santander-Syro, T. Imai, K. Miyamoto, T. Okuda, and D. V. Vyalikh, \textit{Phys. Rev. Lett.} \textbf{124}, 237202 (2020).
%
\end{thebibliography}
\end{document}